\documentclass[sigconf]{acmart}

\usepackage{amsmath}
\usepackage{tcolorbox}
\usepackage{relsize}
\usepackage{hyperref}
\usepackage{balance}

\newcommand{\Qzero}[0]{Do psycholinguistic tests (trained on different text source(s)) reliably infer developer personality from SE communications data?}
\newcommand{\Qone}[0]{RQ1. Do characteristics of the software engineering communications data influence inferred personality?}
\newcommand{\Qtwo}[0]{RQ2. Is one model better than other in inferring developer personality from software engineering data?}
\newcommand{\Qthree}[0]{RQ3. Does the reliability of inferred personality change with the size of the input text?}
\newcommand{\Qfour}[0]{RQ4. Does English proficiency relate to the reliability of personality inferred from psycholinguistic tests?}

\setcopyright{acmcopyright}
\copyrightyear{2021}
\acmYear{2021}
\setcopyright{acmcopyright}\acmConference[ESEM '21]{ACM / IEEE International Symposium on Empirical Software Engineering and Measurement (ESEM)}{October 11--15, 2021}{Bari, Italy}
\acmBooktitle{ACM / IEEE International Symposium on Empirical Software Engineering and Measurement (ESEM) (ESEM '21), October 11--15, 2021, Bari, Italy} \acmPrice{15.00}
\acmDOI{10.1145/3475716.3475775}
\acmISBN{978-1-4503-8665-4/21/10}

\begin{document}

%\title{Infer Personality from GitHub Communications: Promises and Perils}
\title{Promises and Perils of Inferring Personality on GitHub}
%\title{Promises and Perils of Inferring Developer Personality from Communications on GitHub}
%\title{Promises and Perils of Inferring Personality from Software Archives}
%\title{Inferring Personality from Software Repositories: \\Promises and Perils}

\author{Frenk C.J. van Mil}
 \orcid{...}
 \affiliation{Delft University of Technology\\The Netherlands}
 %\email{frenk.van.mil@hotmail.com}
\author{Ayushi Rastogi}
 \orcid{...}
 \affiliation{University of Groningen\\ The Netherlands}
 \email{a.rastogi@rug.nl}
\author{Andy Zaidman}
 \orcid{0000-0003-2413-3935}
 \affiliation{Delft University of Technology\\The Netherlands}
 \email{a.e.zaidman@tudelft.nl}

%\author{Frenk C.J. van Mil, Andy Zaidman, Ayushi Rastogi}
%\orcid{}
%\affiliation{%
%  \institution{Delft University of Technology, the Netherlands}
%}
%  \email{(A.E.Zaidman,a.rastogi)@tudelft.nl}


%The abstract is a short summary of the work to be presented in the article.
\begin{abstract}
\textbf{Background:}
Personality plays a pivotal role in our understanding of human actions and behavior.
Today, the applications of personality are widespread, built on the solutions from psychology to infer personality. 
\textbf{Aim:}
In software engineering, for instance, one widely used solution to infer personality uses textual communication data. 
As studies on personality in software engineering continue to grow, it is imperative to understand the performance of these solutions. 
\textbf{Method:}
This paper compares the inferential ability of three widely studied text-based personality tests against each other and the ground truth on GitHub.
We explore the challenges and potential solutions to improve the inferential ability of personality tests. 
\textbf{Results:}
Our study shows that \emph{solutions for inferring personality are far from being perfect}. 
Software engineering communications data can infer individual developer personality with an average error rate of 41\%. 
In the best case, the error rate can be reduced up to 36\% by following our recommendations\footnote{This work is based on the MSc thesis of Frenk van Mil~\cite{vanMilThesis}. The study data is available at~\cite{Studydata}, while all essentials scripts to replicate our work are available at~\cite{vanMilScripts}. }.

\end{abstract}




\maketitle

\section{Introduction}
%% 1. Importance of studying personality
Personality is an indicator of how we think, feel, and do~\cite{yarkoni2010} with widespread applications. 
In software engineering, for example, individual developer personality is used to understand contribution patterns~\cite{rastogi2016personality}, work preferences \cite{KOSTI2014973}, and work satisfaction \cite{acuna2015}, while collectively, it is used to improve team composition \cite{silva2013, gilal2016}.

\begin{figure}
\includegraphics[width=0.4\textwidth]{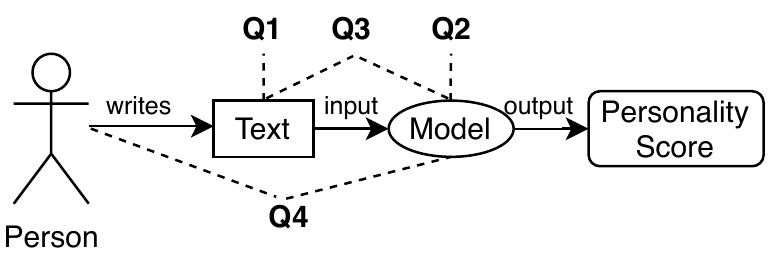}
\caption{shows the working of the psycholinguistic test. \\Q1-4 points to the aspect studied by a research question.}
\label{fig:working}
\end{figure}

% Ways to measure personality
There are two widely used methods to infer personality: questionnaire and psycholinguistic test. 
A questionnaire is a gold standard to measure personality (e.g., \cite{sodiya2007}) in which people are asked a series of questions, responses to which indicate personality.
This approach, however, is time-consuming and relies heavily on the response rate.
On the other hand, a psycholinguistic test fetches a sizeable amount of text written by a person to a `model' to generate personality scores (e.g.,~\cite{rastogi2016personality}; also see Figure~\ref{fig:working}). 

% Identify the gap
Psycholinguistic models are widely used in different contexts (e.g., software engineering~\cite{rastogi2016personality} and social media platforms such as Twitter~\cite{golbeck2011a,golbeck2011b}), often replacing its alternative questionnaire. 
In software engineering, models based on psycholinguistic tests are widely used to measure developer personality~\cite{rastogi2016personality, Calefato:2018:DPL:3196369.3196372}.

While applied to the technical discussions in software engineering, these models are trained on casual conversations (e.g., essays and blog posts~\cite{yarkoni2010}).
Therefore, how well these models infer developer personality is questionable.
%\td{Fill in the blanks}
% Objective of this study
To assess the inferential ability of the models used for measuring developer personality in software engineering, this paper solicits answer to the following research question:

\emph{Do psycholinguistic tests (trained on different text source(s)) reliably infer developer personality from SE communications data?}

% Method
We analyzed developer discussions on collaborative software projects at GitHub using three state-of-the-art and practice psycholinguistic models for inferring personality. 
We studied the Personality Insights tool developed by IBM Watson\footnote{\url{https://www.ibm.com/watson/services/personality-insights/}} and two models from academia: Golbeck et al.~\cite{golbeck2011a} designed for small text sizes such as Twitter posts and 
Yarkoni et al.~\cite{yarkoni2010} designed for longer texts such as blog posts. 
We also generated ground truth by conducting a questionnaire for a subset of developers to validate the personality inferences from the three models.  

We answer the research question in terms of four sub-questions (see Figure~\ref{fig:working} for an overview), exploring the characteristics of 
(1) the input text fetched into a model, 
(2) the model itself, and 
(3)  the person whose personality is measured. 
% Sub-research questions
First, we explored the influence of textual features that do not appear in a usual conversation and are otherwise a part of software engineering communications (e.g., technical jargon) or the syntax used on the platform the discussion ensues (e.g., markdown). 
Our first question is: 

\emph{\Qone}  % Qone defined above.

%Our study shows that ...
%We also provide recommendations on data sanitization steps to improve personality inference from software engineering data. 

%\az{can we be more precise? The reader doesn't know that there are multiple models, so either we explain them before, or we at least mention the number of models that we compare? Coming back to this point, shouldn't we put the paragraph about IBM, Golbeck and Yarkoni earlier?}\fm{I think it would make sense to move the paragraph explaining the models before this one}
Second, we compared models inferring personality to each other and to the ground truth to find: 

\emph{\Qtwo}  % Qtwo defined above.

Next, exploring the relations of the input written data to the model, we investigate:

\emph{\Qthree}  % Qthree defined above.

Finally, linking the characteristics of individuals to that of the model used for inferring personality, we investigate: 

\emph{\Qfour}  % Qfour defined above.

% Contribution
Our study shows that psycholinguistic tests can be administered to infer developer personality from software engineering communications data with an error rate of 25-48\%, with some exceptions.
With our recommended data-sanitization steps, the error rates can be reduced (in our case, up to 36\%). 
We observed that the three models perform comparably and optimally with an average word count between 600 to 1200 and that English proficiency influences the inferred personality traits.

%% 5. Structure of the paper.
%\fm{Should we also list the structure of the paper here?}
This paper is organized into seven sections. 
Starting with an Introduction in Section 1, we describe the background information and related work in Section 2.
Section 3 presents our study design followed by results and their discussion in Section 4 and 5 respectively.
We highlight the limitations and threats to validity of our study in Section 6.
Finally, in Section 7 we present our conclusions and directions for future work.

\section{Background and Related Work}
%% 2. Definition of the Big-Five personality model.
There are many ways to express personality, of which the Big Five Personality model (or BFP) is the most widely used.
The Big Five personality framework has gained recognition among trait psychologists for its validity and reliability \cite{hee2014, mccrae1987}. 
It comprises five traits, namely Openness to Experience, Conscientiousness, Extraversion, Agreeableness, and Neuroticism (in short: OCEAN). 
%Studies show that comparison among people is possible by using the five personality traits' values.
Following is an explanation for the five personality traits:
\begin{itemize}
    \item \textit{Openness to Experience} is characterized by intellectual curiosity, imagination, and open-mindedness. 
It is also referred to as Intellect/Autonomy, or Openness \cite{soto}.
Its opposite, close-minded people often have a narrow range of creativity and intellectual interest. 
    
    \item \textit{Conscientiousness} is characterized by the preference of order, structure, persistence to a goal, and responsibility. 
    Low conscientiousness is linked with comfort, flexibility, and spontaneity but also sloppiness and lack of reliability \cite{soto}.
    \item \textit{Extraversion/Extroversion} is characterized by energy creation from external means, social engagement, and assertiveness. 
    Highly extraverted people feel comfortable in social environments, experience positive emotions more often than introverted people \cite{soto}.
    \item \textit{Agreeableness} is characterized by the general concern for other's well-being and social harmony. 
    Disagreeable people have less concern for the regard of others and social norms of politeness \cite{soto}.
    \item \textit{Neuroticism} is characterized by the tendency to experience negative emotions. 
    Lower neuroticism is linked to emotional stability. Lower neurotic people tend to stay calm and resilient, also referred to as emotional stability \cite{soto}.
\end{itemize}

%Similarities in linguistic characteristics form the basis for personality profiles relative to others with similar characteristics. 
Studies show that the personality structure of an individual is, in part, a function of the linguistic characteristics the individual shares with a group of people \cite{hamilton1957}.
Further, Pennebaker~\cite{pennebaker} adds that the smallest and stealthiest words in our vocabulary define something about our personality; words like `with' and `together' often indicate the author having better social skills, having more friends, and the author usually rate themselves as more outgoing. 
With this as basis, automatic methods called `psycholinguistic models' identify personality from a text by transforming the words used into personality scores.

%% 1. Prior studies. Where is this paper going to build on?
Many studies in software engineering (e.g.,~\cite{rastogi2016personality,calefato2019, 7809476}) and beyond (e.g.,~\cite{golbeck2011a}) have used psycholinguistic tests for inferring personality.
In software engineering, one of the first studies showed
that communications data can be used to infer personality~\cite{rigby2007can}. 
Today, software communications data are used to infer developer personality (individually and as a group) to characterize developers and their participation patterns~\cite{rastogi2016personality,calefato2019}, and its effect on project success~\cite{yun2020personality}, including intermediate steps such as pull request acceptance~\cite{iyer2019effects}.

As the studies exploring the role of personality in software engineering continue to grow, it is crucial to understand the promises and perils of existing solutions and explore future directions, if necessary. 

\section{Study design}
% Introduction to the section.
To investigate the inferential ability of psycholinguistic tests on software engineering communications, we compare the test scores from three state-of-the-practice models to each other and the ground truth. 
This section presents how we infer developer personality using psycholinguistic tests and gauge ground truth using a questionnaire.
Next, we describe the data collected for analysis followed by the statistical tests to investigate each research question. 
A curated list of data and scripts used for analysis (in compliance with the GDPR) is available at respectively~\cite{Studydata} (data) and~\cite{vanMilScripts} (scripts).

\subsection{Psycholinguistic Tests}
At the core of psycholinguistic tests are models that take written text as input and transform it to generate five numbers, representing the five personality traits (see Figure~\ref{fig:working}).
In this study, we use three widely used models: two from academia (Yarkoni~\cite{yarkoni2010} and Golbeck et al. \cite{golbeck2011a, golbeck2011b}) and one from industry (IBM) - Personality Insights.\footnote{https://www.ibm.com/watson/services/personality-insights/}

% How do the two academic models work?
The two academic models are based on Linguistic Inquiry and Word Count tool (LIWC)\footnote{http://liwc.wpengine.com/}.
LIWC calculates the percentage of words in a text indicating emotions and part of speech, among others~\cite{tausczik2010} (also referred to as \emph{word categories}). 
These word categories are correlated with personality traits; therefore, calculating a weighted sum of word categories and correlation coefficient indicates personality.

% Difference between the two models
While the two models are structurally the same, they are different in the text sources for training.
Yarkoni trained its model on long essays~\cite{yarkoni2010}, while Golbeck trained it on short Tweets~\cite{golbeck2011a, golbeck2011b}.
These differences in text culminates into correlations, hence differences in the weights of the two models. 
%\ar{Can we say more about the nature of text?}\az{Good point from Ayushi, explain in 1-2 sentences how this text is likely different from Github data}\fm{Differences: Markdown, SE related terms, more formal compared to Tweets, prescribed structures for e.g., PRs?}
Likewise, we can expect differences to culminate in software engineering text which is different from tweets and essays due to the use of markdown, SE-specific terms, formality, and structure of pull requests.

% Explain how PI works.
Unlike the two academic models, Personality Insights (or PI) is not trained on one type of text source and is expected to be more general-purpose compared to the two academic models.
It uses an open vocabulary and a machine learning model that continues to learn new words, phrases, topics, and categories~\cite{schwartz2013}.
This model acts as a black box generating personality scores for a given text.

% Explain how we used transformation and normalization on scores.
\emph{Preprocessing.} 
The three models generate five real numbers each, indicating the five personality traits. 
However, these numbers do not have a meaning in themselves and show personality relative to a population.
For example, 
imagine two people, Alice and Bob, with an extraversion score of 1.25 and -2.3. 
How do we interpret the two scores? 
How big is the extraversion score 1.25 compared to the extraversion score -2.3? 
%\fm{Reacting to Andy's comment below, I think we could make more clear what our intentions are with this paragraph if we take two fictive people. On with a score of 1.25 for extraversion and one with -2.3.}

To make the inferred personality scores meaningful, we bring them to a comparable scale, assuming that the sample population represents all personality types. 
We transform the numeric scores for each separate personality trait using min-max normalization to values including and between 0 and 1. 
In the revised scale, extraversion score 0 refers to an introvert, relative to the studied population.
Similarly, score 1 refers to an extrovert, and a score of 0.5 refers to an average personality trait with combined introvert and extrovert characteristics.
%\az{Reacting to your first comment here... I think the 2 paragraphs together are quite essential, i.e., the normalisation helps interpretation. Should we also come up with a fictive persona that we classify in terms of personality to make it easier to grasp?}

\subsection{Questionnaire}
% Choice of method for ground truth
We use personality traits inferred from a questionnaire (also a gold standard) as our ground truth. 
Currently, there are two widely used questionnaires for inferring personality: NEO-PI-R \cite{costa2008revised} and Big Five Inventory (BFI)~\cite{john1999big, john1991, john2008}.
We choose to use BFI driven by two factors. 
One, we intended to reach a wider audience since only then can we reliably interpret the inferential ability of psycholinguistic tests. 
%\az{How does BFI relate to the wider audience? For your second reason the length of the questionnaire is the explaining factor (Very good! But what about reason 1?} \fm{We believe that a shorter questionnaire will lead to more people reacting. Might need a source.}
Two, we realize that people are less likely to react to our questionnaire since answers to our questions indicate their personality, and they may have privacy concerns.
BFI is the shorter of the two methods, as filling in the questionnaire takes 5-10 minutes, compared to the 30-40 minutes required for the NEO-PI-R questionnaire. Furthermore, BFI's reported reliability and validity are comparable to NEO-PI-R~\cite{fossati2011, ALANSARI2016S209, Arterberry2014}.
We hypothesize that by selecting a less time-intensive questionnaire, we can reduce the number of participants dropping out, enabling us to reach a wider audience.
We further incentivized participation by offering an Amazon gift card of 25 USD as prize. 
To mitigate privacy concerns, we informed our prospective survey respondents on our objective, implementation details, and their right to withdraw, in compliance with GDPR\footnote{https://gdpr-info.eu/} and as approved by the University Ethics Board via Data Privacy Impact Assessment.

% Survey description
Our survey comprises 44 questions indicating personality (Openness (10), Extraversion (8), Agreeableness (9), Conscientiousness (9), and Neuroticism (8)). 
An example question in our survey is: `I am someone who is talkative.'
The answer to each question is a value between 1 and 5, with 1 representing `strongly disagree' and 5 - `strongly agree.'

Personality is then calculated as a sum of the answers to the following question numbers for a given personality trait~\cite{john2008,john1991, benet1998} (see the survey in replication package for detail).
\begin{itemize}
 \item Openness: 5, 10, 15, 20, 25, 30, 35R, 40, 41R, 44
    \item Conscientiousness: 3, 8R, 13, 18R, 23R, 28, 33, 38, 43R
    \item Extraversion: 1, 6R, 11, 16, 21R, 26, 31R, 36
    \item Agreeableness: 2R, 7, 12R, 17, 22, 27R, 32, 37R, 42
    \item Neuroticism: 4, 9R, 14, 19, 24R, 29, 34R, 39
\end{itemize}
here, `R' is a reversed score calculated as 6 - <score> .

In addition to the questions relating to personality, we also asked questions indicating proficiency in English: (1) `English is my mother tongue' (2) `I am fluent in written English', and (3) `In what country did you spend most your youth?'. The first two questions accepted a three-point Likert scale (`yes,' `no,' and `maybe') as valid answers.
We also informed the participants about the compliance of our investigation to the GDPR.

% \begin{figure}[ht]
%     \centering
%     \includegraphics[width=\linewidth]{Images/continent_distribution_responses.png}
%     \caption{Distribution of responses to the questionnaire per continent.}
%     \label{fig:content_distribution}
% \end{figure}
% \fm{We could use this figure in the story. Might need to mention somewhere the over-representation of Europe.}

\subsection{Data Collection}
% Overview
We select contributors to infer personality by mining their communication history relating to software development and can gather ground truth by running a questionnaire. 
For some contributors, we cannot infer personality.
This includes contributors who did not communicate on GitHub, or their communication history was insufficient or unavailable.
%\az{The phrasing of the previous sentence implicitly says that for some contributors, we cannot infer personality. If this is your intention, then we need to explain!} \fm{We could not do it for people for whom we had no or less than 100 words (as PI would not work)} 
We analyze development activities on GitHub for our analysis~\cite{KalliamvakouMSR2014}, as is also used by the recent studies on personality (e.g.,~\cite{rastogi2016personality, iyer2019effects}).  

% What are our project selection criteria?
We collected information on the top 3\% of projects with most development activities from GHTorrent \cite{Gousi13} (version June 2019). 
We set a lower bound of 33 pull requests for project selection (same as the new dataset for pull request research~\cite{zhang2020shoulders}) since such projects are less likely to exhibit communicative intentions amongst developers. %\az{Previous sentence... there are 2 arguments here: the strict development process and the communicative intentions. I would not make the strict development process explicitly, simply because we really don't know. A project with 4 pull requests might be super-structured... the key thing that we are after is communication, so let's stick to that} 
Projects deleted at the time of creating the dataset were left out, as we could not trace the comments back to GitHub anymore. 
Ultimately, we collected data from 8,436 projects developed in Java, JavaScript, Python, Ruby, Go, and Scala. 
%\ar{What is the original count we started with?} \fm{I don't know this number. It was in my mailbox, but I don't have access to it anymore. We might need to check with Xunhui for this.} \az{Maybe the original numbers are not that important, but rather the number of projects that these comments originate from?} \fm{@Andy, that's 8,436 projects. Or do you mean something different?}
The selected projects may have more contributors. However, we leave out contributors who did not write any comments or whose comments are not available for analysis.

%  how did we transform communications to personality?
For each selected contributor, we had at least one written comment (in a pull request, issue, or commit) and possibly many comments available in the selected projects.
While each contributor comment can suggest personality, we combine all available sentences and present them as the input to the three models for improved reliability. %\az{I think it is very good that at this point you are specifying that you are looking at PR comments, because there are other forms of communication on GH as well, think issues. Perhaps we should be even more clear about this earlier on in the paper?} \fm{We did not make a distinction between PR, issue, or commit comments. In other words, we used (almost) all possible comments on GH.}

\emph{Preprocessing.} 
Before fetching the comments' data to the model, we manually analyzed the comments and found two outlier behavior.
First, some contributors had written less than 100 words, even after combining the texts from all written comments.
Since shorter texts are less likely to reliably infer personality (e.g., PI tool denies request with less than 100 words~\cite{ibm_docs}), we removed contributors with shorter texts from our analysis. 
Second, we found some accounts with an extraordinarily large size of text (e.g., \textit{lintr-bot}, a bot for static code analysis for R).
Upon closer inspection of these comments, we found that texts in such accounts have repetitive statements and texts such as \texttt{``I am a bot''} suggesting that these accounts are not operated by humans but by bots.
We manually identified and after inspection removed all such accounts based on keyword search `bot' and otherwise large text size.
Finally, we analyze 4,081,957 comments written by 28,337 contributors from 8,436 projects.

% Explain how we selected people for the ground truth.
\emph{Survey.} From the 28,337 contributors, we chose a representative sub-sample worldwide, accounting for the observed regional differences in personality traits~\cite{kajonius2017cross, schmitt2007}.
Using K-Means clustering, we created six clusters on the world map, one centroid for each continent (except Antarctica).
The location information for these contributors is inferred using the Bing Map API service\footnote{\url{https://www.bingmapsportal.com/Application}}, in combination with the information available on GHTorrent~\cite{Gousi13}.
We randomly selected 2,050 participants from the six clusters, equally from each cluster for whom we could infer email addresses.
We invited 2,050 participants to answer our research question. %\az{Should we further say why we think that a worldwide representation is important?} \fm{Think so. We might back this up with the out-commented section I pasted below:}
% Possibly this difference is introduced by demographics or culture. For example, earlier studies found lower openness scores for people in East-Asia than for people in Europe \cite{kajonius2017cross, schmitt2007}. Mak and Tran \cite{MAK2001181} found for Asians with a high English proficiency (in terms of fluency) a positive correlation with openness values. In certain cultures, adjectives, generally used as a major support in taxonomic approaches, does not cover the domain sufficiently to allow for identification \cite{rolland2002}. In our ground-truth, however, we do not observe a significant difference in means between Europe ($M=0.70$) and Asia ($M=0.69$), and America ($M=0.73$) and Asia nor do we find any significant differences in the inferred scores of the methods between continents. We do, however, not have information about the culture and demographics of our participants to prove their influence.

% How we tried boosting our survey response rate?
To boost the survey response rate, we send customized emails to contributors at 10:00 AM in their timezone \cite{kent2004}.
In the end, 267 people filled our questionnaire (a 13\% response rate).
A detailed description of the demographics of our survey respondents is presented in Table~\ref{tab:continent_answers}. 
Our participants over-represent Europe and the USA, but we have a representation of each continent. 

%distribution of people from all around the globe (see Figure \ref{fig:world_map}). 

%\begin{figure}[ht]
%    \centering
%    \includegraphics[width=\linewidth]{Images/Map_shadow_full.png}
%    \caption{All identified users from GitHub, obtained from the GHTorrent dataset, mapped roughly on the world map.}
%    \label{fig:world_map}
%\end{figure}

\begin{table}[]
\caption{presents the distribution of survey responses continent-wise and English proficiency inferred from self-perceived fluency and mother tongue English.}
\label{tab:continent_answers}
\begin{smaller}
\begin{tabular}{|l|c|c|c|c|c|c|}
\hline
\rowcolor[HTML]{656565} 
{\color[HTML]{FFFFFF} } &
  \multicolumn{3}{c|}{\cellcolor[HTML]{656565}{\color[HTML]{FFFFFF} Fluent}} &
  \multicolumn{3}{c|}{\cellcolor[HTML]{656565}{\color[HTML]{FFFFFF} Mother tongue}} \\ \hline
\rowcolor[HTML]{656565} 
{\color[HTML]{FFFFFF} } &
  {\color[HTML]{FFFFFF} Yes} &
  {\color[HTML]{FFFFFF} No} &
  {\color[HTML]{FFFFFF} Maybe} &
  {\color[HTML]{FFFFFF} Yes} &
  {\color[HTML]{FFFFFF} No} &
  {\color[HTML]{FFFFFF} Maybe} \\ \hline
\cellcolor[HTML]{656565}{\color[HTML]{FFFFFF} Europe}        & 102 & 7 & 11 & 8  & 110 & 2 \\ \hline
\rowcolor[HTML]{EFEFEF} 
\cellcolor[HTML]{656565}{\color[HTML]{FFFFFF} Asia}          & 49  & 9 & 9  & 5  & 59  & 3 \\ \hline
\cellcolor[HTML]{656565}{\color[HTML]{FFFFFF} North-America} & 40  & 0 & 0  & 34 & 6   & 0 \\ \hline
\rowcolor[HTML]{EFEFEF} 
\cellcolor[HTML]{656565}{\color[HTML]{FFFFFF} South-America} & 15  & 2 & 1  & 0  & 18  & 0 \\ \hline
\cellcolor[HTML]{656565}{\color[HTML]{FFFFFF} Africa}        & 12  & 0 & 2  & 5  & 8   & 1 \\ \hline
\rowcolor[HTML]{EFEFEF} 
\cellcolor[HTML]{656565}{\color[HTML]{FFFFFF} Total} &
  {\color[HTML]{000000} 218} &
  {\color[HTML]{000000} 18} &
  {\color[HTML]{000000} 23} &
  {\color[HTML]{000000} 52} &
  {\color[HTML]{000000} 201} &
  {\color[HTML]{000000} 6} \\ \hline
\end{tabular}
\end{smaller}
\end{table}

% What statistical tests do we use?
\subsection{Statistical tests}
To answer the four questions, we need information on:
(1)~do there exist differences in the personality scores calculated using two methods?
(2) if a difference exists, how big is the effect? and
(3) overall accuracy of a method.

% How to show if there exist differences between the two models?
To check whether there exists a statistically significant difference in the personality scores inferred using two methods, we use the Student t-test~\cite{student} or the Wilcoxon signed-rank test~\cite{wilcoxon}.
When the data is (near) normally distributed, we use the Student t-test and Wilcoxon signed-rank test otherwise.
We use a combination of the Shapiro-Wilk test and Q-Q plots to infer the non-normality of the data. 

% How do we measure effect size?
If there are statistically significant differences in the distribution of personality scores inferred using any two models, we calculate effect size to indicate the actual differences in personality scores.
We report effect size r \cite{rosenthal1994parametric}, Cohen's d \cite{cohen2013statistical, cohen1992statistical}, or Cram\'{e}rs V \cite{cramer1999mathematical} depending on data distribution and choice of method (see Table \ref{tab:cohen_interpretation} for interpreting effect sizes). 
Cohen's d is used for normally distributed data, and we use r otherwise~\cite{cohen2013statistical, cohen1992statistical}.
Cram\'{e}rs V is used for nominal data.

%\ar{When do we use each of the three techniques is not evident in the write-up?}
%\ar{Cite the sources for interpretation in Table 1} \fm{Interpretations are according to \cite{cohen2013statistical, cohen1992statistical}} \fm{More info in Thesis Chapter 3.4 page 15}


% How do we express accuracy?
Finally, we report the accuracy of a method compared to the ground truth.
We report accuracy in Mean Absolute Error (MAE) and Root Mean Squared Error (RMSE). 
MAE considers each error equally, and by calculating an average of the errors, it gives an average impression of the overall error. 
RMSE, in contrast, takes the square root of the average squared error, which means that more significant errors count more than minor errors.
Collectively, the two metrics present an overall and nuanced view of errors with values closer to zero, indicating similarity in scores. 
In this paper, we report MAE and both MAE and RMSE in the technical report~\cite{vanMilThesis}.
%\ar{What is the difference between the two techniques? Why do we report the two values? Do we report different values in different contexts?}
 All statistical tests are conducted in R, details on which are available in the replication package.
 
 \begin{table}[!t]
\caption{presents effect size and its interpretation}
\label{tab:cohen_interpretation}
\begin{tabular}{llll}
\hline
\rowcolor[HTML]{656565} 
{\color[HTML]{FFFFFF} Effect size} & {\color[HTML]{FFFFFF} d}           & {\color[HTML]{FFFFFF} r}           & {\color[HTML]{FFFFFF} Cramér's V}   \\ \hline
\multicolumn{1}{|l|}{negligible} & \multicolumn{1}{l|}{\textless 0.2} & \multicolumn{1}{l|}{\textless 0.1} & \multicolumn{1}{l|}{\textless 0.07} \\ \hline
\multicolumn{1}{|l|}{small}        & \multicolumn{1}{l|}{\textless 0.5} & \multicolumn{1}{l|}{\textless 0.3} & \multicolumn{1}{l|}{\textless 0.21} \\ \hline
\multicolumn{1}{|l|}{medium}       & \multicolumn{1}{l|}{\textless 0.8} & \multicolumn{1}{l|}{\textless 0.5} & \multicolumn{1}{l|}{\textless 0.35} \\ \hline
\multicolumn{1}{|l|}{large}        & \multicolumn{1}{l|}{$\ge$ 0.8}        & \multicolumn{1}{l|}{$\ge$ 0.5}        & \multicolumn{1}{l|}{$\ge 0.35$}        \\ \hline
\end{tabular}
\end{table}

% %%%%%%
% For each question, what did we do exactly.
% %%%%%%
\subsection{Characteristics of SE communications data}
\emph{\Qone}\\
% Describe the problem
Psycholinguistic tests for inferring personality are trained on natural language text.
When applied to GitHub communications data, the written text has two other sources of variability: 
(1) technical words relating to software engineering, and 
(2) GitHub flavored language (i.e., markdown markup language\footnote{https://daringfireball.net/projects/markdown/}).
So, suppose the model proposed by Golbeck et al. were to infer personality. 
It can incorrectly take exclamation mark (used to integrate images as \texttt{![...](...)}) as an indicator for high conscientiousness, high neuroticism, and low openness to experience~\cite{golbeck2011a}.

% how did we identify the preprocessing steps?
To systematically identify factors that can potentially influence the inferred personality score, we looked into four directions.
First, we identified preprocessing steps used by the previous studies in software engineering~\cite{DBLP:journals/corr/ArnouxXBMAS17, schoonvelde2019, SAGADEVAN2015201, alamsyah2019, carducci2018}.
Second, we searched for terms in LIWC dictionary that can be incorrectly represented in software engineering text. 
We looked for extreme/outlier personality scores to identify such elements and propose preprocessing steps based on our findings.
Third, we investigated the influence of elements that raises privacy concerns (e.g., emails and IP addresses - European GDPR \cite{opinion_gdpr, eu-679-2016}).
Finally, we investigate the effect of removing elements for improved efficiency and reduced storage requirements.

% Which preprocessing steps did we use?
We identified twelve preprocessing steps indicating (a) the influence of software engineering on the written text, (b) platform-specific features (e.g., Markdown), and (c) identifying/personal information. %\fm{This sentence feels redundant with the previous paragraph.}
We identified removing (1) numbers~\cite{DBLP:journals/corr/ArnouxXBMAS17, schoonvelde2019, SAGADEVAN2015201, alamsyah2019}, 
(2) hashtag~\cite{carducci2018, DBLP:journals/corr/ArnouxXBMAS17},
(3) URLs~\cite{SAGADEVAN2015201, carducci2018, alamsyah2019, DBLP:journals/corr/ArnouxXBMAS17}, and 
(4) @-references~\cite{alamsyah2019} from previous studies.
Relating to the terms incorrectly represented in software engineering, we identified removing (5) quotes, 
(6) code blocks, and 
(7) images as part of Markdown characteristics.
For privacy reasons, we explore removing (8) email addresses and (9) IP addresses.
Finally, for efficiency, we propose removing (10) upper case, (11) variants of white space (e.g., \textbackslash r, \textbackslash n, and \textbackslash t) replacing it with a single white space, and (12) double white space and space before punctuation. 
Some or all of these steps can potentially influence the inferential ability of the psycholinguistic models for inferring personality in software engineering. 

% How did we find if a preprocessing step was good?
\emph{Approach.} To gauge the impact of a preprocessing step, we compute personality scores with and without the step. 
%\fm{'factor' may not be clear if we used 'preprocessing step' earlier}
If we find statistically significant differences in the distribution of scores (for paired observations), we explore whether the scores are better with or without the preprocessing step. 
Next, we compare the scores to the ground truth to see if the proposed change is for the better.
We report the accuracy of the inferred personality scores using Mean Absolute Error and Root Mean Squared Error.

\subsection{Comparison of models}
\emph{\Qtwo}\\
% How did we approach this question?
We compare the scores inferred from each model to the ground truth and report accuracy in MAE and RMSE.
But before we answer this question, we process the data in two ways.
One, we applied the appropriate preprocessing steps identified in the previous research question to the input written text.
Two, we manually investigate the distribution of scores in the three models to identify any need for transformation to bring the data from the three models to the same scale.
We apply mean-centering such that the new mean for each score is zero \cite{Iacobucci2016}. 
Finally, we compare the scores inferred from each model (original and mean-centered) to the ground truth.

%%%%%%%%%%%%%%%%%%%% COPY PASTE %%%%%%%%%%%%%%%%%%%%%%%%%%%%%%%%%%%%

\subsection{Size of input text}
\emph{\Qthree}\\
% Why this research question?
Golbeck is trained on Twitter messages of size 50 to 5724 words~\cite{golbeck2011a}, while Yarkoni is trained on blog posts with at least 50,000 words~\cite{yarkoni2010}.
The text size mentioned above refers to the concatenation of available Twitter messages and blog posts for an author. 
%\az{In the above sentence, I think it is important to mention that this concerns concatenated Twitter messages/blog posts...}
PI requires all text to have at least a hundred words and allows up to an estimated 42,000 words\footnote{PI allows for JSON formatted input of $250KB=250*1024B=256,000B$ in size. ASCII encoding uses 8 bits (=1 byte) per character. If we take an average of 5.1 letters per English word~\cite{wolframalpha}, with one space after each word, this gives an estimated $256,000/(5.1+1) \approx 42,000$ words.}.
%\ar{describe the footnote in detail}

% How did we compare different text sizes?
To identify the optimal text size for inferring personality scores, we compare subsets of the same text (100, 600, 1200, and 3000 words) to each other.
Our choice of sizes is inspired by PIs analysis of optimal text size \cite{ibm_docs}.
In our dataset, we have 4,346 contributors who had written at least 3000 words are candidate for analysis. 
% How did we compare significance?
We select the first `n' words for each text size to identify differences in personality score inferred using different text sizes. 
If vast differences exist, we compare the accuracy of the increased text size to the ground truth.

\subsection{English proficiency}
\emph{\Qfour}\\
% Definitions for proficiency.
We explore English proficiency in two ways.
Our first definition links \emph{English mother tongue}, often referred to as the child's native or first acquired language \cite{mizza2014}, to English proficiency.
Second, we explore \emph{fluency}, referring to one's ability to express oneself easily, as an indication of English proficiency.
The first definition is objective and captures the sub-population who innately speak English.
The second definition is somewhat subjective but caters to the sub-population who acquired English proficiency by other means. 

% How did we assess proficiency?
Survey responses to the question on English proficiency classified contributors into three groups: (1) English proficiency -yes, (2) English proficiency - no, and (3) English proficiency - maybe.
We left the option `maybe' since it showed considerable overlap with the yes and no groups, as inferred from the English Proficiency Index.~\footnote{https://www.ef.com/wwen/epi/}
% Comparison of both definitions of proficiency.
We compared the personality scores for the groups (yes and no) to each other to identify if there exist differences in inferred personality. 
We use unpaired tests for this question (i.e., Wilcoxon summed rank test and unpaired t-test).
If there exist differences, we calculate its extent (by comparing it to ground truth) to find the influence of English proficiency on inferred personality.
%\fm{In the questionnaire we also allow for 'maybe' as answer. Should we mention that we left this group out, as it showed much overlap with both the yes and no groups?}


\section{Results}
%% 1. Introduction of the section.

%% 2. Answering of research question 1.
 % Qone defined above.
\begin{tcolorbox}
\textbf{\Qone}\\
\emph{Eliminating platform-specific features (e.g., quotes and code block) improved inferred personality scores. 
Other factors did not affect personality scores but otherwise improved efficiency (e.g., double white space) and lowered privacy concerns (e.g., email address).}
\end{tcolorbox}

Our investigation of twelve preprocessing steps shows either improvement in the inferred personality score or improvements in efficiency and reduced privacy concerns without influencing the inferred personality score. 
Table~\ref{tab:preprocessing_improvements} presents the maximum percentage improvements in the inferred personality traits (calculated as mean absolute error) when a preprocessing step is applied. 
The reported scores represent the maximum for the five personality traits (OCEAN) applied on the three models.
Table~\ref{tab:preprocessing_improvements} also shows the percentage of the text population that is affected by the preprocessing step.
Remember that the characteristics of the text do not change with the choice of model or personality trait.
We report findings from 11 out of 12 preprocessing steps.
We do not present results from removing quoted text since this text is written by someone else. 
We do not consider the quoted text as a representation of the analyzed person's personality.

\begin{table}[]
\caption{Reports maximum percentage improvement in MAE and population affected for all traits for a preprocessing.}
%\caption{For each preprocessing step the percentage maximum improvement in MAE found for all traits and the maximum affected population. The removal of quotes is not listed, as the quotes cannot be part of the analyzed person's personality.}
\label{tab:preprocessing_improvements}
\begin{tabular}{lll}
\rowcolor[HTML]{656565} 
{\color[HTML]{FFFFFF} \begin{tabular}[c]{@{}l@{}}Preprocessing  step\end{tabular}} &
  {\color[HTML]{FFFFFF} \begin{tabular}[c]{@{}l@{}}Improvement MAE (\%)\end{tabular}} &
  {\color[HTML]{FFFFFF} \begin{tabular}[c]{@{}l@{}}Population  (\%)\end{tabular}} \\ \hline
\rowcolor[HTML]{FFFFFF} 
\multicolumn{1}{|l|}{\cellcolor[HTML]{FFFFFF}Code blocks} &
  \multicolumn{1}{l|}{\cellcolor[HTML]{FFFFFF}\textless{}36\%} &
  \multicolumn{1}{l|}{\cellcolor[HTML]{FFFFFF}\textless{}100\%} \\ \hline
  \rowcolor[HTML]{EFEFEF} 
  \multicolumn{1}{|l|}{\cellcolor[HTML]{EFEFEF}Remove URLs} &
  \multicolumn{1}{l|}{\cellcolor[HTML]{EFEFEF}\textless{}18.2\%} &
  \multicolumn{1}{l|}{\cellcolor[HTML]{EFEFEF}\textless{}99.9\%} \\ \hline
\rowcolor[HTML]{FFFFFF} 
\multicolumn{1}{|l|}{\cellcolor[HTML]{FFFFFF}Remove images} &
  \multicolumn{1}{l|}{\cellcolor[HTML]{FFFFFF}\textless{}11.1\%} &
  \multicolumn{1}{l|}{\cellcolor[HTML]{FFFFFF}\textless{}99\%} \\ \hline
  \rowcolor[HTML]{EFEFEF} 
  \multicolumn{1}{|l|}{\cellcolor[HTML]{EFEFEF}Remove numbers} &
  \multicolumn{1}{l|}{\cellcolor[HTML]{EFEFEF}\textless{}9.1\%} &
  \multicolumn{1}{l|}{\cellcolor[HTML]{EFEFEF}\textless{}99.9\%} \\ \hline
\rowcolor[HTML]{FFFFFF} 
  \multicolumn{1}{|l|}{\cellcolor[HTML]{FFFFFF}Remove @-ref} &
  \multicolumn{1}{l|}{\cellcolor[HTML]{FFFFFF}\textless{}6.7\%} &
  \multicolumn{1}{l|}{\cellcolor[HTML]{FFFFFF}\textless{}93.9\%} \\ \hline
  \rowcolor[HTML]{EFEFEF} 
 \multicolumn{1}{|l|}{\cellcolor[HTML]{EFEFEF}Remove IP} &
  \multicolumn{1}{l|}{\cellcolor[HTML]{EFEFEF}\textless{}5.3\%} &
  \multicolumn{1}{l|}{\cellcolor[HTML]{EFEFEF}\textless{}95.5\%} \\ \hline
 \rowcolor[HTML]{FFFFFF}     
  \multicolumn{1}{|l|}{\cellcolor[HTML]{FFFFFF}Remove hashtags} &
  \multicolumn{1}{l|}{\cellcolor[HTML]{FFFFFF}\textless{}5\%} &
  \multicolumn{1}{l|}{\cellcolor[HTML]{FFFFFF}\textless{}91.4\%} \\ \hline
   \rowcolor[HTML]{EFEFEF}   
\multicolumn{1}{|l|}{\cellcolor[HTML]{EFEFEF}Lowercase parsing} &
  \multicolumn{1}{l|}{\cellcolor[HTML]{EFEFEF}0\%} &
  \multicolumn{1}{l|}{\cellcolor[HTML]{EFEFEF}\textless{}0.1\%} \\ \hline
 \rowcolor[HTML]{FFFFFF}
\multicolumn{1}{|l|}{\cellcolor[HTML]{FFFFFF}Whitespace parsing} &
  \multicolumn{1}{l|}{\cellcolor[HTML]{FFFFFF}0\%} &
  \multicolumn{1}{l|}{\cellcolor[HTML]{FFFFFF}0\%} \\ \hline
\rowcolor[HTML]{EFEFEF} 
\multicolumn{1}{|l|}{\cellcolor[HTML]{EFEFEF}Remove emails} &
  \multicolumn{1}{l|}{\cellcolor[HTML]{EFEFEF}0\%} &
  \multicolumn{1}{l|}{\cellcolor[HTML]{EFEFEF}\textless{}0.1\%} \\ \hline
\rowcolor[HTML]{FFFFFF} 
\multicolumn{1}{|l|}{\cellcolor[HTML]{FFFFFF}Remove spaces} &
  \multicolumn{1}{l|}{\cellcolor[HTML]{FFFFFF}0\%} &
  \multicolumn{1}{l|}{\cellcolor[HTML]{FFFFFF}0\%} \\ \hline
\end{tabular}
\end{table}

We found that removing code blocks, quotes, images, URLs, and numbers improves the models' accuracy. 
The first three factors -- code blocks, quotes, images -- are platform and software engineering-specific features, and as expected, their removal improves inferred personality scores.
We recommend to always remove quotes and code blocks as they can reflect another person's personality.
Likewise, by removing the markdown text indicative of images, which otherwise can be linked with personality, we reduce the chances of misclassification.
Removing URLs also improved our inference, a factor identified in prior studies~\cite{SAGADEVAN2015201, carducci2018, alamsyah2019, DBLP:journals/corr/ArnouxXBMAS17}.
We believe that the improvement is attributed to the words in the URL which are otherwise misclassified as a part of the communication.

Another factor identified in the existing studies is the removal of numbers~\cite{DBLP:journals/corr/ArnouxXBMAS17, schoonvelde2019, SAGADEVAN2015201, alamsyah2019}.
Generally, our analyses show that removing numbers improves the inferred personality scores, except for Yarkoni's extraversion and agreeableness.
For Yarkoni's extraversion and agreeableness, we observed that removing numbers made the inference less accurate. 
See the technical report for details~\cite{vanMilThesis}. %\fm{Should we here add a reference to the anonymous link you just created?}.
On further investigating Yarkoni's model, we found that the word category \emph{Number} correlates to extraversion and agreeableness~\cite{yarkoni2010}, and hence the observation.
`Number' has no relation to any other personality trait in the other two models. Golbeck does not use it and there is no effect on PI.

Other preprocessing steps did not influence the personality score but improved another property.
We found that removing the upper casing and spaces before punctuation and conformance of white space did not influence inferred personality (refer to Table~\ref{tab:preprocessing_improvements}) but improved processing speed and reduced storage needs.
The remaining factors: hashtags~\cite{carducci2018, DBLP:journals/corr/ArnouxXBMAS17} and @-references/usernames~\cite{alamsyah2019} (identified in the literature) and email addresses and IP addresses (found in this study), did not influence personality score but reduced privacy concerns by eliminating personally identifiable information.

For the remainder of this study, we applied all twelve preprocessing steps identified above on the input text by removing them from the text.
The only exceptions are Yarkoni's extraversion and agreeableness for which we retained the word category `number'.


%% 3. Answering of research question 2.
 % Qtwo defined above.
\begin{tcolorbox}
\textbf{\Qtwo}\\
\emph{The three models perform comparably when brought to the same scale.}
\end{tcolorbox}

\begin{figure*}
    \centering
    \includegraphics[width=0.85\linewidth]{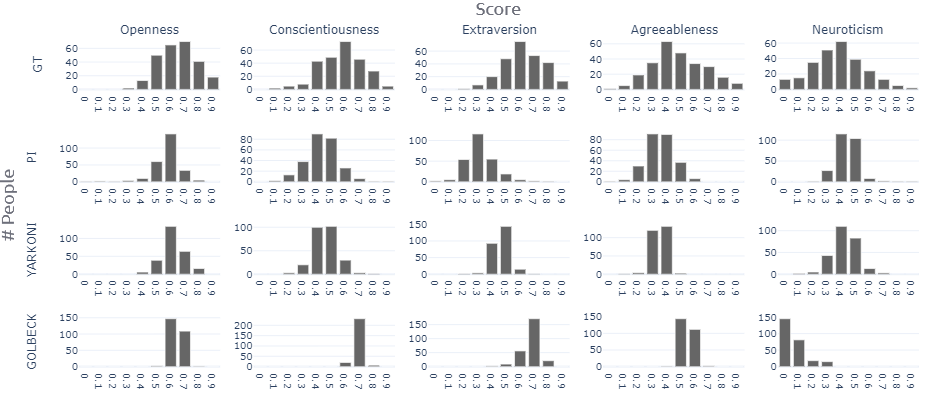}
    \caption{presents the personality traits distribution for the three psycholinguistic tests. Row 1 shows the ground truth (GT) in comparison to PI, Yarkoni and Golbeck. Each histogram presents the distribution of personality scores for a single trait.
}
    \label{fig:data_full_hist}
\end{figure*}

\begin{figure}[ht]
    \centering
    \includegraphics[width=0.55\textwidth]{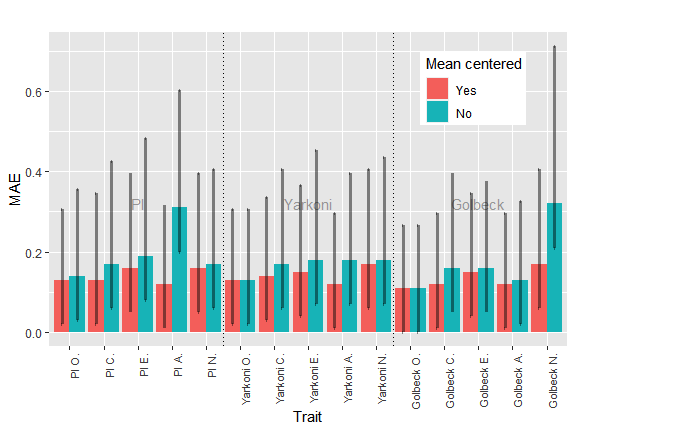}
    \caption{MAE scores for each personality trait with and without mean-centering}
    \label{fig:mae_mean_centering}
\end{figure}

%As we found in RQ2, the models have an error rate of 30-40\% with 95\% confidence. This raised the question if the inaccuracy is caused by the person analyzed. We suspected the fluency and mother tongue of a person to have a possible influence on the outcome. It is well established that word recognition in the second language can be affected by the native language \cite{lemhofer2008}, leading to the use of different words and phrases. With the use of different words, this also means that the psycholinguistic models may give different results. From our results, we did find that all methods show lower openness scores non-fluent people.

Figure \ref{fig:data_full_hist} presents the distribution of personality scores for each of the three models and the ground truth.
Figure \ref{fig:data_full_hist} shows that PI and Yarkoni have a similar distribution compared to Golbeck.
Generally, Golbeck has a high average personality score indicating big outliers, except for neuroticism.

Figure \ref{fig:data_full_hist} also indicates a similar distribution in personality scores, albeit at different scales.
Therefore, before comparing any two models, we mean-centered the scores.
We observed that while the differences among the three methods are still statistically significant, the differences in MAE and RMSE decrease to have a negligible effect when mean-centered.
The differences among the models with and without mean-centering is shown in Figure~\ref{fig:mae_mean_centering}.
This suggests that the three models perform comparably but at different scales. 
A detailed comparison of the differences among the models with and without mean centering is available in the technical report~\cite{vanMilThesis}.
%\ar{Can there be a table indicating the differences between models once mean centered?} \fm{Instead of a table, Figure~\ref{fig:mae_mean_centering} is added.}

%\ar{Is there a table, similar to the previous question, that presents comparison of the results of the three models?}\az{Good point. If this is meant to be Table 6, we need to do a better job at explaining how to interpret it!} \fm{A table with the effects of the transformations and the effect for each these models? Or just a comparison between the numbers of the three methods?}

\begin{table}[ht]
\setlength{\tabcolsep}{2.8pt}
\caption{Maximum absolute error in inferring personality scores at 90\%, 95\%, and 99\% confidence intervals.}
\label{tab:maximum_absolute_error}
\begin{smaller}
\begin{tabular}{|c|c|c|c|c|c|c|c|c|c|c|c|c|c|c|c|}
\hline
\rowcolor[HTML]{656565} 
{\color[HTML]{FFFFFF} } &
  \multicolumn{5}{c|}{\cellcolor[HTML]{656565}{\color[HTML]{FFFFFF} PI}} &
  \multicolumn{5}{c|}{\cellcolor[HTML]{656565}{\color[HTML]{FFFFFF} Yarkoni}} &
  \multicolumn{5}{c|}{\cellcolor[HTML]{656565}{\color[HTML]{FFFFFF} Golbeck}} \\ \hline
\rowcolor[HTML]{656565} 
{\color[HTML]{FFFFFF} } &
  {\color[HTML]{FFFFFF} O} &
  {\color[HTML]{FFFFFF} C} &
  {\color[HTML]{FFFFFF} E} &
  {\color[HTML]{FFFFFF} A} &
  {\color[HTML]{FFFFFF} N} &
  {\color[HTML]{FFFFFF} O} &
  {\color[HTML]{FFFFFF} C} &
  {\color[HTML]{FFFFFF} E} &
  {\color[HTML]{FFFFFF} A} &
  {\color[HTML]{FFFFFF} N} &
  {\color[HTML]{FFFFFF} O} &
  {\color[HTML]{FFFFFF} C} &
  {\color[HTML]{FFFFFF} E} &
  {\color[HTML]{FFFFFF} A} &
  {\color[HTML]{FFFFFF} N} \\ \hline
\multicolumn{1}{|l|}{\cellcolor[HTML]{656565}{\color[HTML]{FFFFFF} 90\%}} &
  \multicolumn{1}{l|}{.28} &
  \multicolumn{1}{l|}{.36} &
  \multicolumn{1}{l|}{.39} &
  \multicolumn{1}{l|}{.51} &
  \multicolumn{1}{l|}{.36} &
  \multicolumn{1}{l|}{.25} &
  \multicolumn{1}{l|}{.32} &
  \multicolumn{1}{l|}{.41} &
  \multicolumn{1}{l|}{.40} &
  \multicolumn{1}{l|}{.37} &
  \multicolumn{1}{l|}{.22} &
  \multicolumn{1}{l|}{.31} &
  \multicolumn{1}{l|}{.32} &
  \multicolumn{1}{l|}{.26} &
  \multicolumn{1}{l|}{.61} \\ \hline
\rowcolor[HTML]{FFFFFF} 
\cellcolor[HTML]{656565}{\color[HTML]{FFFFFF} 95\%} &
  .34 &
  .43 &
  .46 &
  .54 &
  .41 &
  .29 &
  .38 &
  .48 &
  .45 &
  .43 &
  .25 &
  .38 &
  .37 &
  .31 &
  .68 \\ \hline
\rowcolor[HTML]{EFEFEF} 
\cellcolor[HTML]{656565}{\color[HTML]{FFFFFF} 99\%} &
  .42 &
  .53 &
  .55 &
  .61 &
  .50 &
  .38 &
  .46 &
  .56 &
  .52 &
  .50 &
  .29 &
  .51 &
  .44 &
  .42 &
  .77 \\ \hline
\end{tabular}
\end{smaller}
\end{table}

When on the same scale, we found that most traits can be inferred with a 25-48\% error rate from the ground truth with 95\% confidence, except Golbeck neuroticism (68\%) and PI agreeableness (54\%) (see Table~\ref{tab:maximum_absolute_error}).
Table~\ref{tab:maximum_absolute_error} also shows error rates at 90\% and 99\% confidence. 
With error margins this high, one must be cautious in interpreting personality scores, particularly with the two extreme cases: PI agreeableness and Golbeck neuroticism. 
For these two traits, the high error rate renders the results meaningless.

When inferring individual personality, in the worst case, this implies that a person deemed high on agreeableness can actually be low on agreeableness. 
This problem, however, can somewhat subside when seen collectively (as is in the case of group or team personality).
In case of group personality, aggregation of scores can reduce the effect of error margins contingent on the choice of aggregation technique (e.g., median instead of mean). This group personality are the personal characteristics or qualities shared by the members of the group, and the group personality composition has been observed to influence group effectiveness in domains outside of software engineering~\cite{Halfhill2005}.
We believe that personality inferred from the three psycholinguistic tests will be a better representation of the team than individuals.

Other than the above, we also observed that transformations reduced the effect of large outliers, improving the scores. 
For future research, we recommend applying transformation(s) to minimize the effects of outliers. 

%% 4. Answering of research question 3.
% Qthree defined above.
\begin{tcolorbox}
\textbf{\Qthree}\\ 
\emph{Inferential ability of PI and Yarkoni is optimal at 600-1200 words and 600 words for Golbeck. 
Text with less than 100 words gives unreliable estimates, while beyond 3000 words, we expect no further improvements.}
\end{tcolorbox}

For each model, we compared personality scores with increasing text size: 100, 600, 1200, and 3000 (similar to the research of PI \cite{ibm_docs}). 
We compared personality inferences of text size 100 to text size 600, 600 to 1200, and 1200 to 3000.
The optimal text size is the one after which there are no significant improvements in the inferred personality scores with the increasing text size.
Table~\ref{tab:word_change_table} presents the optimal text size for each personality trait of the three models.
Generally, the inferred personality scores increased with text size.
However, for Yarkoni conscientiousness and Golbeck neuroticism, we did not find a consistent pattern and hence no optimal text size.
 
 \begin{table}[ht]
\caption{Optimal text size for personality inferences}
\label{tab:word_change_table}
\begin{smaller}
\begin{tabular}{|l|l|l|l|}
\hline
\rowcolor[HTML]{656565} 
{\color[HTML]{FFFFFF} } & {\color[HTML]{FFFFFF} PI} & {\color[HTML]{FFFFFF} Yarkoni} & {\color[HTML]{FFFFFF} Golbeck} \\ \hline
\cellcolor[HTML]{656565}{\color[HTML]{FFFFFF} O} & 600  & 100  & 3000 \\ \hline
\rowcolor[HTML]{EFEFEF} 
\cellcolor[HTML]{656565}{\color[HTML]{FFFFFF} C} & 1200 & ?    & 600  \\ \hline
\cellcolor[HTML]{656565}{\color[HTML]{FFFFFF} E} & 1200 & 1200 & 600  \\ \hline
\rowcolor[HTML]{EFEFEF} 
\cellcolor[HTML]{656565}{\color[HTML]{FFFFFF} A} & 3000 & 3000 & 600  \\ \hline
\cellcolor[HTML]{656565}{\color[HTML]{FFFFFF} N} & 1200 & 600  & ?    \\ \hline
\end{tabular}
\end{smaller}
\end{table}

Our comparison of personality scores inferred with text sizes 100, 600, 1200, and 3000 words show that an optimal text size for PI and Yarkoni mostly ranges from 600 to 1200 words.
After 1200 words, we did not see significant improvements in personality score. 
For Golbeck, the optimal text size is 600 words.
This is a lower word count compared to the previous estimates by PI~\cite{ibm_docs} and Yarkoni~\cite{yarkoni2010} that shows no significant improvements in MAE after 3000 words.
Golbeck does not provide such an estimate, but we have no reason to believe it needs more than 3000 words. 
%\ar{We may not have sufficient data.}\az{I also think this claim is a bit unfounded... even it you weaken it by saying ``we believe''. What is the basis for this believe?}

%% 5. Answering of research question 4.
%Qfour defined above.
\begin{tcolorbox}
\textbf{\Qfour}\\ 
\emph{Fluency in English influences openness to experience scores. 
Other traits are inferred more accurately for non-fluent people by PI and Yarkoni and by Golbeck for fluent people. }
\end{tcolorbox}

\begin{table}[]
\caption{presents traits for which inferred personality changes with the choice of model and its effect size.}
\label{tab:effect_size_proficiency}
\begin{smaller}
\begin{tabular}{|c|c|c|c|c|c|c|}
\hline
\rowcolor[HTML]{656565} 
{\color[HTML]{FFFFFF} } &
  \multicolumn{3}{c|}{\cellcolor[HTML]{656565}{\color[HTML]{FFFFFF} Fluency}} &
  \multicolumn{3}{c|}{\cellcolor[HTML]{656565}{\color[HTML]{FFFFFF} Mother tongue}} \\ \hline
\rowcolor[HTML]{656565} 
{\color[HTML]{FFFFFF} } &
  {\color[HTML]{FFFFFF} } &
  \multicolumn{2}{c|}{\cellcolor[HTML]{656565}{\color[HTML]{FFFFFF} MAE}} &
  {\color[HTML]{FFFFFF} } &
  \multicolumn{2}{c|}{\cellcolor[HTML]{656565}{\color[HTML]{FFFFFF} MAE}} \\ \hline
\rowcolor[HTML]{656565} 
{\color[HTML]{FFFFFF} Trait} &
  {\color[HTML]{FFFFFF} \begin{tabular}[c]{@{}c@{}}Difference\\ effect size\end{tabular}} &
  {\color[HTML]{FFFFFF} Yes} &
  {\color[HTML]{FFFFFF} No} &
  {\color[HTML]{FFFFFF} \begin{tabular}[c]{@{}c@{}}Difference\\ effect size\end{tabular}} &
  {\color[HTML]{FFFFFF} Yes} &
  {\color[HTML]{FFFFFF} No} \\ \hline
\rowcolor[HTML]{FFFFFF} 
PI A      & small  & .32 & .23 & -      &  -   &    - \\ \hline
\rowcolor[HTML]{EFEFEF} 
Yarkoni N & medium & .18 & .13 & -      & -    & -    \\ \hline
\rowcolor[HTML]{FFFFFF} 
Golbeck O & large  & .13 & .16 & medium & .13 & .11 \\ \hline
\end{tabular}
\end{smaller}
\end{table}

We observed that, depending on the model, English proficiency is linked to the differences in personality traits.
Generally, PI and Yarkoni generate somewhat less accurate scores for fluent people in comparison to the ground truth.
Golbeck, on the contrary, generates less accurate scores for non-fluent people (refer to the technical report for all the scores~\cite{vanMilThesis}). 
Specifically, PI agreeableness and Yarkoni neuroticism present worse scores for people fluent in English by a small and medium amount, respectively (see Table~\ref{tab:effect_size_proficiency}).
In the case of Golbeck openness to experience, we observe significant differences relating to fluency and medium differences for mother tongue in favor of fluent people.
%\ar{Can we make Table 6 also show if it makes it worse for fluent and English mother tongue?} \fm{Done}

\begin{table}[ht]
\caption{Mean openness scores for each method and the ground truth for fluent and non-fluent people and people with and without English as their mother tongue.}
\label{tab:openness_all_methods}
\begin{smaller}
\begin{tabular}{l|l|l|l|l|}
\cline{2-5}
 &
  \multicolumn{2}{l|}{\cellcolor[HTML]{656565}{\color[HTML]{FFFFFF} Fluent}} &
  \multicolumn{2}{l|}{\cellcolor[HTML]{656565}{\color[HTML]{FFFFFF} Mother tongue}} \\ \hline
\rowcolor[HTML]{656565} 
\multicolumn{1}{|l|}{\cellcolor[HTML]{656565}{\color[HTML]{FFFFFF} Method}} &
  {\color[HTML]{FFFFFF} Yes} &
  {\color[HTML]{FFFFFF} No} &
  {\color[HTML]{FFFFFF} Yes} &
  {\color[HTML]{FFFFFF} No} \\ \hline
\rowcolor[HTML]{FFFFFF} 
\multicolumn{1}{|l|}{\cellcolor[HTML]{FFFFFF}PI}           & 0.63 & 0.6  & 0.63 & 0.63 \\ \hline
\rowcolor[HTML]{EFEFEF} 
\multicolumn{1}{|l|}{\cellcolor[HTML]{EFEFEF}Yarkoni}      & 0.68 & 0.64 & 0.68 & 0.67 \\ \hline
\rowcolor[HTML]{FFFFFF} 
\multicolumn{1}{|l|}{\cellcolor[HTML]{FFFFFF}Golbeck}      & 0.7  & 0.67 & 0.71 & 0.7  \\ \hline
\rowcolor[HTML]{EFEFEF} 
\multicolumn{1}{|l|}{\cellcolor[HTML]{EFEFEF}Ground truth} & 0.71 & 0.64 & 0.74 & 0.7  \\ \hline
\end{tabular}
\end{smaller}
\end{table}

% O is affected in all
That said, openness to experience generally changes with fluency for all the three models, with less fluent people being inferred as less open to experience (see Table~\ref{tab:openness_all_methods} for details).
We observed a similar pattern for the ground truth.
The mean score of fluent people ($M=0.71$) is significantly lower than the mean of non-fluent people ($M=0.65$), $V=6.55$, $p<0.05$. This could indicate that the differences found for openness are introduced by the population, not by the methods used.
In addition, mother tongue English is linked to higher openness to experience than the sub-population whose mother tongue is not English, except for PI, which shows the same (refer Table~\ref{tab:openness_all_methods}).

This finding in itself is not surprising, as several studies have shown the link of cultural background~\cite{rolland2002} and geographical location~\cite{schmitt2007} to the openness to experience, which can be linked to English proficiency.
Earlier studies found lower openness scores for the people in East-Asia than for the people in Europe \cite{kajonius2017cross, schmitt2007}. 
Mak and Tran \cite{MAK2001181} further added that Asians with high English proficiency (in terms of fluency) are reportedly more open to experience. 
 
Our ground-truth, however, did not show a significant difference in the means among Europe ($M=0.70$), Asia ($M=0.69$) and America ($M=0.73$).
This also applied to the three methods, which did not significantly differ in the inferred personality scores between continents. 
We do not have information about the culture and demographics of our participants to study its influence.

%\az{Can we take anything actionable from this analysis?}
Our findings imply that when inferring personality using psycholinguistic tests, we can incorrectly infer personality scores for two reasons: differences in the background of the participants (including English proficiency) and the limitations of the model itself.

%However, if we look into the openness scores obtained with the questionnaire, we can already observe lower average openness scores for non-fluent ($M=0.65$) compared to fluent people ($M=0.71$), $V=6.55$, $p<.05$. Possibly this difference is introduced by demographics or culture. Possibly this difference is introduced by demographics or culture. For example, earlier studies found lower openness scores for people in East-Asia, compared to people in Europe~\cite{kajonius2017cross, schmitt2007}. Mak and Tran~\cite{MAK2001181} found for Asians with a high English proficiency (in terms of fluency) a positive correlation with openness values. In certain cultures, adjectives, generally used as a major support in taxonomic approaches, does not cover the domain sufficiently to allow for identification \cite{rolland2002}. In our ground-truth, however, we do not observe a significant difference in means between Europe ($M=0.70$) and Asia ($M=0.69$), and America ($M=0.73$) and Asia nor do we find any significant differences in the inferred scores of the methods between continents. We do, however, not have information about the culture and demographics of our participants to prove their influence. Future studies could investigate the influence of culture and demographics on psycholinguistic models.

\section{Discussion and Implications}
%In this section, we put our results into context answering the research question.
%We present how to use these findings and the extend to which these results are useful.
%Finally, we discuss the limitations of our study and future directions.

\begin{tcolorbox}
\textbf{\Qzero} 
\emph{Partially yes}
\end{tcolorbox}
% Crux of our study
Our study shows that irrespective of the choice of psycholinguistic test (with different text size requirements) and the application of proposed preprocessing steps, personality traits can be inferred with an average error rate of 41\% at 95\% confidence.
While we found that twelve out of the thirteen preprocessing steps can improve personality inference up to 36\% (in the best case), personality traits such as Golbeck neuroticism can have an error rate up to 68\%. 

% How does it link to the literature?
With error margins this high, individual personality inferences are far from an accurate depiction.
Notably, the high error rates found here corroborate with the existing research inferring personality using psycholinguistic tests in software engineering~\cite{rastogi2016personality}, but also broadly~\cite{kramer2014experimental}. 
Other than the limits of the psycholinguistic tests, our study further highlights the role of English proficiency.
The current personality traits make an implicit assumption that the written text is only a reflection of author. 
In reality, proficiency in English - an individual characteristic, can influence the written text and hence the inferred score.

With such fault margins, we urge people to be careful with concluding from the inferred personality scores. Therefore, a peril of all covered proposed psycholinguistic methods, currently, is their ability to predict all personality types accurately for individuals.
A such, we should be very much aware that the inferred personality can be very different from the actual personality. Therefore, we issue a stark warning that this approach should not be used for making decisions relating to individuals. In the case of personality inference on an individual level, one must be careful to use the scores as an indicator, not a truth value. While considering the possible error, the personality scores can be effectively used for team formations or group-related research.
However, we do expect that the approach can work better when analyzing the personality at a group level. 
When personality is measured for a group, depending on the choice of aggregation techniques (e.g., median instead of mean), psycholinguistic models can offer a reasonable estimate. 

%Nonetheless, the error margins are high enough to incorrectly classify developer personality...
%It affects x\% of the population... 
Next, we present recommendations on choosing a model and optimally inferring developer personality from software communications data.

\begin{tcolorbox}
\textbf{What model should I choose?}
\emph{Any of the three}
\end{tcolorbox}

Our study shows that no one model is better than others in inferring developer personality from software engineering data, except when looking for specific personality traits.
Irrespective of the choice of text sources (e.g., tweets vs. blogs), or its count (single data source vs. multiple data sources), the personality inferred from the three models are similar, with some exceptions. 
We recommend that with an appropriate amount of text size, all the models perform similarly.
This is particularly helpful now that PI is deprecated
\footnote{https://cloud.ibm.com/docs/personality-insights?topic=personality-insights-about\#about} 
- a widely used option recently.

\begin{tcolorbox}
\textbf{How to infer developer personality optimally?}
\emph{Clean the data, choose any of the three models with optimal text size and interpret the findings in relation to author's English proficiency and known error margins.}
\end{tcolorbox}

We recommend a three-step process starting with preprocessing the software communications data based on the 12 steps identified in this study.
Next, apply any model or choose a model that reportedly works best for a specific personality trait of interest. 
For instance, studies interested in personality trait neuroticism should avoid Golbeck's model (see Table~\ref{tab:maximum_absolute_error}). 
Depending on the choice of model, choose a minimum text size for optimally reliable scores. 
For example, when using Golbeck, 600 words will suffice while use 600-1200 words for Yarkoni.
Ultimately, once the personality scores are found, it is essential to interpret them within the specific context of the person whose traits are inferred and the error margins when using psycholinguistic tests.

% Introduction to the next question
This brings us to the next question that while we can infer developer personality from software communications data,
\begin{tcolorbox}
\textbf{... should we apply psycholinguistic tests to software engineering data to infer developer personality?}
\emph{Yes and No}
\end{tcolorbox}
%\fm{I changed the above box to make it clearer that the question in the box belongs to the sentence above. Please check if you like this change.}

We will answer this question through two sub-questions: 
(1) Can we apply? and
(2) Should we apply?
The answer to the first question is yes.
We can apply psycholinguistic tests to infer developer personality at scale, but we need to be aware of the \% error rates.
%\ar{How do we come up with such a number?} \fm{I would suggest sticking with the 30-40\% error rate. So "...be aware that personality scores may have a 30-40\% error rate.' Or something similar}
Despite the error rates found here, many studies in software engineering have demonstrated the applications of inferring developer personality. 
Recent studies in software engineering have shown how developer personality can reflect in their contributions~\cite{rastogi2016personality, calefato2019}, pull request acceptance~\cite{iyer2019effects} and project success~\cite{yun2020personality}.

The second aspect of this question is `should we apply?', and the answer here is tricky.
On the one hand, studies on personality have improved our understanding of development practices and software development, in general.
The other extreme is the perpetual harm that studies of this kind can bring to an individual and community.
For instance, filtering candidates for hiring merely based on personality traits can not only lead to false conclusions but also discrimination against certain personality types~\cite{stone2005personality}.
Another factor is mental health.
Studies show that bipolar disorder can influence personality inference~\cite{chang2016subconscious}, since these inferences are a snapshot in time.
Alternatively, a wrong judgement on personality (e.g., during hiring process) can have a backlash on the mental health.

%\az{This is an important statement, and a necessary one. We could reinforce it by giving a concrete case?} \fm{Companies should not deny people based on personality. Ethics section 9.3.2 of the Thesis. And mental health section 9.3.3 of the Thesis.}

% Ethics (?)
%\emph{Interpretation}
%With all the results we found, it is important to note the possible threat the misuse of results can be. As shown in RQ2, the scores for people can be inaccurate. Personality inference can be wrong in multiple ways: the context, but also the score itself could be misunderstood. In no way should personality inference be used as a tool \textit{on its own} to accept or reject people for job applications. Filtering candidates merely on personality traits might not only lead to false conclusions but also discrimination on personality types \cite{stone-romero2005}.

%All in all, even though the act of personality inference is seemingly possible, one should always feel obliged to honor the human behind the numbers and mind the possible negative impact the numbers may induce.

\begin{tcolorbox}
\textbf{Can psycholinguistic tests perform better on software engineering text?}
\emph{Maybe}
\end{tcolorbox}

This question too can be divided into two parts: 
(1) Can existing models perform better? and
(2) Can we design optimal psycholinguistic models specific to software engineering?

We have no reasons to believe how and why existing models can perform better.
We used models used in academia and industry; these models are trained on different sized text (small for tweets and long for essays), different sources (single vs. multiple), and trained once vs. constant learning.
Despite these variabilities in the model, we did not find any model performing better than the other.  
Therefore, we believe that existing models cannot perform better.  
One possibility can be to rethink how we separate natural text from software engineering text, as suggested by Bachelli et al.~\cite{bacchelli2012content}.

That said, we can design psycholinguistic models specific to software engineering.
The solutions proposed in this study optimize personality inference on syntactic elements such as removing markdown.
Software communications data, however, have semantics that work differently in software engineering than a usual conversation.
For example, a term such as `cookies' is assigned to the word category `bio,' which has a special meaning in the software engineering context.
The term `cookie' can also mean the food we understand from everyday life, but less likely in a software engineering context.
With the current research, it is not evident whether designing an optimal psycholinguistic model specific to software engineering will help.
More work is required to substantiate the claim.

\subsection{Implications}
%Our research has implications for research, practice, and education.
\emph{Research:} Our study presents the promises and perils of mining GitHub communications data for inferring developer personality.
These findings can serve as a guideline for future research building on psycholinguistic tests for understanding a software engineering phenomenon.
Further, by highlighting the limits of psycholinguistic tests for inferring personality in software engineering, our study opens up avenue for next steps (see Section 7 for future work).

\emph{Practice:} Our study shows the practical usability of the existing solutions and the ethical concerns that one should consider prior to use.
Further, our study offers a frame of reference on (1) how to infer personality scores optimally (based on the existing psycholinguistic models)? and (2) how to interpret it?

\emph{Education:} Our study shows how signals (in our case derived from software communications data) can infer complex concepts, such as personality and means of doing it. 

\subsection{Comparison to related work}
If we consider earlier studies on personality among software engineers, the found scores are not unexpected. In the study by Calefato et al.~\cite{Calefato:2018:DPL:3196369.3196372} the mean openness found among their participating developers (M = 0.79) is reasonably close to the mean openness found for this study (M = 0.71). Similarly, their mean for conscientiousness (M = 0.6) and agreeableness (M = 0.64) are close to our means for conscientiousness (M = 0.63) and agreeableness (M = 0.68). As earlier studies show similar distributions of scores, this indicates that our extraction process for personality traits from text works reasonably well. 

In terms of the accuracy of automatically extracting personality traits from text, we compare our work against the large-scale analysis of personality in software engineering by Calefato et al.~\cite{calefato2019}. In their comparison of personality models, they observe an accuracy ranging from 40--70\%, which is globally in line with our own results.

\section{Limitations and Threats to Validity}
%% Limitations found within the project:
%In the below, we discuss the potential consequences of our design choices and the limits of our work. 

\noindent\textbf{Construct validity.~}% Design choices
Our analysis is as good as the ground truth.
We used a lightweight questionnaire - which is the closest we had to the gold standard with scalability.
Another choice we made relates to the written text. 
We selected the first `n' number of words written by a person to gauge their personality. 
Had we chosen the middle `n' words or the last `n' words, our findings could have been different, but mostly since studies show that personality evolves over time, although slowly \cite{rastogi2016, Calefato:2018:DPL:3196369.3196372}.
Another related factor is the size of the text. 
We selected contributors whose written text (at least 100 words) is available for analysis. 
This, on the one hand, defines the limit of our approach. On the other hand, it can systematically exclude some personality traits.
The same argument applies to surveys, where the respondents may have self-selection bias.

\smallskip
\noindent\textbf{Internal validity.~} 
Our study presents inferred personality scores by applying preprocessing steps.
While we systematically identify these steps, we might have missed steps that do not fall in our purview (e.g., SHA references or emojis). 
These words do not contribute to personality scores but add to word count, thereby likely influencing the inference. 
Another factor that can potentially inflate the reported percentage improvements is the normalization of personality scores.
While we normalize the data to help understand the personality scores, change in a person's personality score after preprocessing can influence the normalized scores of others.
This is a necessary trade-off, but we advise our readers to remember this potential side-effect while gauging the potential of preprocessing steps. 

\smallskip
\noindent\textbf{External validity.~}% Generalizability
Finally, the usefulness of our results is as good as the sub-populations to which it apply.
We have no reasons to believe why and how software communications data on other platforms will differ from Github (other than the differences in the features of the platform), suggesting that our findings should be generalizable.
Also, we believe that the factors found important here should apply to other platforms, although their relative relevance can change.
We will need more studies to substantiate these claims.

To counter self-selection, we have performed random undersampling on the majority class to ensure that each continent is equally represented. Through this sampling, we select 2,050 participants accounting for the observed regional differences in personality traits~\cite{kajonius2017cross, schmitt2007}.

\section{Conclusions and Future Work}
This paper comes as a guideline for inferring personality using software engineering communications data.
By comparing the personality scores inferred from three state-of-the-practice models to the ground truth, we provide recommendations on the promises and perils of mining GitHub communications data for inferring personality scores.
We identify 12 preprocessing steps that improve personality inferences, yet the average error rate is 41\% with 95\% confidence.
We also identify optimal text size for reliable personality inferences and recommend choosing any of the three models with some exceptions.
Finally, we highlight the role of English proficiency and error margins while interpreting personality scores. 

Knowing the limits of the existing solutions, future research should take one of the two possible directions.
One, propose a solution specific to software engineering (e.g., process the software engineering text to resemble natural conversations or modify the model to perform optimally on software engineering communications data).
Alternatively, look for personality cues in places other than text (e.g., software code and activity patterns) as indicators of personality.

\begin{acks}
We thank all survey participants and Xunhui Zhang for technical support. This research was partially funded by the Dutch science foundation NWO through the Vici ``TestShift'' grant (No. VI.C.182.032).
\end{acks}

\balance
\bibliographystyle{ACM-Reference-Format}
\bibliography{infer_personality}

\end{document}